\documentstyle[12pt,epsf]{article}
\def\href#1#2{#2}   
\newif\ifdraft
\draftfalse	  
\reversemarginpar   
\makeatletter
\let\mlabel=\label
\let\adkendequation=\endequation%
\def\endequation{\adkendequation\adklabel\global\@ignoretrue}
\let\adkendeqnarray=\endeqnarray%
\def\endeqnarray{\adkendeqnarray\adklabel\global\@ignoretrue}
\newbox\marglabbox
\def\adklabel{\ifvoid\marglabbox\else\marginpar{\unhbox\marglabbox}\fi}
\def\label#1{\ifdraft\ifmmode%
  \global\setbox\marglabbox=\hbox{\hfill\fbox{\tiny\verb*~#1~}}%
  \else\ifinner\else\marginpar{\hfill\fbox{\tiny\verb*~#1~}}%
  \fi\fi\fi \mlabel{#1}}
\makeatother
\ifdraft%
\fi
\font\twelvebb=msbm12
\font\tenbb=msbm10
\font\sevenbb=msbm7
  \newfam\bbfam
  \def\bb{\fam\bbfam\twelvebb}
  \textfont\bbfam=\twelvebb
  \scriptfont\bbfam=\tenbb
  \scriptscriptfont\bbfam=\sevenbb
\font\twelveeusm=eusm10 scaled 1200
\font\teneusm=eusm10
  \newfam\eusmfam
  
  \textfont\eusmfam=\twelveeusm
  \scriptfont\eusmfam=\teneusm
  \scriptscriptfont\eusmfam=\scriptfont\eusmfam
\font\twelvefrak=eufm10 scaled 1200
\font\tenfrak=eufm10
  \newfam\frakfam
  \def\frak{\fam\frakfam\twelvefrak}
  \textfont\frakfam=\twelvefrak
  \scriptfont\frakfam=\tenfrak
  \scriptscriptfont\frakfam=\scriptfont\frakfam
\def\sqr#1#2{{\vcenter{\hrule height.#2pt
   \hbox{\vrule width.#2pt height#1pt \kern#1pt
      \vrule width.#2pt}
   \hrule height.#2pt}}}

\def\bsqr#1#2{{\vrule width #1pt height#2pt}}
\def\bsquare{{\mathchoice\bsqr66\bsqr66\bsqr33\bsqr33}}
\def\badbreak{\penalty1000}
\newtheorem{theorem}{Theorem}
\newtheorem{lemma}{Lemma}
\newtheorem{definition}{Definition}
\newtheorem{corollary}{Corollary}

\newtheorem{proposition}{Proposition}
\newenvironment{proof}{{\em Proof.}}{\badbreak$\;\bsquare$\smallskip}
\def\deg{\mathop{\rm deg}}		    
\def\identity{{\bb I}}			    
\def\union{\cup}                            
\def\intersection{\cap}                     
\def\ord{\mathop{\rm ord}}                  
\def\C{{\bb C}}				    
\newcommand{\Dset}{{\frak D}}               
\newcommand{\D}{{\bf D}}                    
\newcommand{\Dr}{{\overline D}}             
\newcommand{\G}{{\bf G}}                    
\newcommand{\Gr}{{\overline G}}             
\newcommand{\Fr}{{\overline F}}             
\newcommand{\Gsl}{{{\frak G}^{sl}}}         
\newcommand{\Hcal}{{\cal H}}                
\newcommand{\psibar}{\overline{\psi}}       
\newcommand{\pbar}{\overline{p}}            
\newcommand{\gmu}{\gamma_{\mu}}             
\newcommand{\gnu}{\gamma_{\nu}}             
\newcommand{\dmunu}{\delta_{\mu,\nu}}       
\newcommand{\gone}{\gamma_{1}}              
\newcommand{\gtwo}{\gamma_{2}}              
\newcommand{\gfive}{\gamma_{5}}             
\newcommand{\oA}{\overline{A}}              
\newcommand{\oB}{\overline{B}}              
\newcommand{\oC}{\overline{C}}              
\newcommand{\oQ}{\overline{Q}}              
\newcommand{\cC}{{\cal C}}                  
\newcommand{\cZ}{{\cal Z}}                  
\newcommand{\cH}{{\cal H}}                  
\newcommand{\bfa}{{\mathbf a}}              

\begin{document}

\title{
  Strong Non-Ultralocality of Ginsparg-Wilson \\
  Fermionic Actions}

\author{Ivan Horv\'ath\thanks{{\tt ih3p@virginia.edu}} \\[1ex]
  Department of Physics, University of Virginia\\
  Charlottesville, Virginia 22903, U.S.A. \\[1ex]
  Chetan~T~Balwe\thanks{{\tt cbalwe@umich.edu}} \\[1ex]
  Department of Mathematics, University of Michigan\\
  Ann Arbor, Michigan  48109, U.S.A. \\[1ex]
  Robert Mendris\thanks{{\tt mendris@math.ohio-state.edu}} \\[1ex]
  Department of Mathematics, Ohio State University\\
  Columbus, Ohio 43210, U.S.A.}

\date{June 29 2000}

\maketitle

\begin{abstract}
  \noindent
  It is shown that it is impossible to construct a free theory of 
  fermions on infinite hypercubic Euclidean lattice in even number
  of dimensions that: (a) is ultralocal, (b) respects the symmetries 
  of hypercubic lattice, (c) chirally nonsymmetric part of its 
  propagator is local, and (d) describes single species of massless 
  Dirac fermions in the continuum limit. This establishes 
  non-ultralocality for arbitrary doubler-free Ginsparg--Wilson fermionic 
  action with hypercubic symmetries 
  {\it (``strong non-ultralocality'')}, 
  and complements the earlier general result on non-ultralocality of 
  infinitesimal Ginsparg--Wilson--L\"uscher symmetry transformations
  {\it (``weak non-ultralocality'')}.
\end{abstract}
   
\section{Introduction}

Proper inclusion of chiral dynamics in the framework of 
lattice-regularized gauge theory has long been one of the main themes 
in this nonperturbative approach. While there are several useful angles 
to look at this, the many years of conceptual confusion were primarily
caused by the simple fact that lattice theories with naive chiral 
symmetry can not reproduce the physics of chiral 
anomaly~\cite{Kar81A,Nie81A}. At the same time, the departure from naive 
chiral symmetry to accommodate anomalies typically mutilates other aspects 
of chiral dynamics.

However, the set of acceptable lattice fermionic actions is rather large, 
and so it is not implausible that there are actions without naive chiral
symmetry, where both the anomalies and the chiral dynamics are accommodated
simultaneously. Guided by renormalization group arguments, Ginsparg and 
Wilson suggested that one should examine lattice Dirac theories in which 
the chirally nonsymmetric part of the massless propagator is {\it local} 
(coupling decays at least exponentially at long distance)~\cite{Gin82A}. 
This is very plausible since the important aspects of chiral dynamics 
are typically associated with the long distance behavior of the 
propagator, which is then very mildly affected by the chirally 
nonsymmetric part of the action. Lattice fermionic actions with this 
property became 
known as Ginsparg--Wilson (GW) actions and vectorlike theories 
defined by them indeed have all the important ingredients of chiral 
dynamics~\cite{Gin82A, Has98B, Cha98A}. More importantly, otherwise 
acceptable GW actions can be explicitly constructed as demonstrated by 
the class of operators of Neuberger's type~\cite{Neu98BA, Her98A}. 
 
Formally appealing feature of the GW approach is that it can be defined
through the symmetry principle~\cite{Lus98A}. The underlying 
Ginsparg--Wilson--L\"uscher (GWL) symmetry operation is rather 
unconventional and depends on the action itself. It has been proved
that if the action is symmetric, then the corresponding infinitesimal
GWL transformation can not be {\it ultralocal}~\cite{Hor99A}. In other 
words, the transformation must rearrange the system in such a way that 
a given variable is mixed with infinitely many other variables on an 
unbounded lattice. This is a universal property of GWL symmetry 
in the presence of hypercubic symmetries 
{\it (weak non-ultralocality)}, and it shows that to enforce 
features of chiral dynamics without naive 
chiral symmetry requires a delicate cooperation of many degrees of 
freedom. Aside from conceptual value, weak non-ultralocality has a well 
defined physical meaning, because it is associated with 
non-ultralocality of Noether currents corresponding to GWL symmetry. 

While understanding general structural properties of GW actions is of 
prime importance for both conceptual and practical reasons, the
progress has been rather slow. This is mainly due to the fact that
the defining condition for GW actions is nonlinear, which causes 
the related problems to be quite nontrivial. An important basic question
here is whether GW actions can be ultralocal. This issue has been 
settled (with negative answer) for commonly studied class of GW 
actions some time ago~\cite{Hor98A}, and the most general result published 
so far actually follows from weak non-ultralocality. In particular, it was 
shown that all GW actions for which chirally nonsymmetric part of the 
propagator is ultralocal (and nonzero), are non-ultralocal in the 
presence of hypercubic symmetries~\cite{Hor99A}.

The purpose of this paper is to demonstrate {\it strong non-ultralocality}
as formulated in Ref.~\cite{Hor99A}, i.e. non-ultralocality of all 
doubler-free GW fermionic actions respecting the symmetries of the 
hypercubic lattice. Our proof will follow the strategy developed in 
Ref.~\cite{Hor99C},\footnote{An attempt to prove some new
non-ultralocality result using ideas of~\cite{Hor99C} was also 
announced in~\cite{Bie99A}. However, as pointed out in Ref.~\cite{Hor00A}, 
the claims there do not appear to be substantiated.}
which suggests that it is sufficient to consider two-dimensional momentum
restrictions for free theories and investigate the consequences
of the GW property at the origin of the Brillouin zone.
This leads to an algebraic problem reflecting factorization properties of 
certain class of polynomials~\cite{Hor99C}. We will adopt
the general formalism of Ref.~\cite{Hor99A} and develop some basic notions 
of algebraic geometry that will be needed to keep the presentation
self-contained.

\section{Lattice Dirac Operators and Ultralocality}

Consider an infinite $d$-dimensional hypercubic lattice, where $d$ 
is an even integer. Let $\psi,\psibar^T$ be the vectors of 
fermionic variables associated with lattice sites. We will be 
interested in quadratic expressions (actions) $\psibar \G\psi$, defined
by the linear operator $\G$ which acts in the corresponding linear space.
The usual gauge--flavor structure of the linear space will be ignored 
since it will be sufficient for our purposes to consider only operators 
in unit gauge background. Consequently, the fermionic variables have 
$2^{d/2}$ spinorial components, and arbitrary matrix $\G$ can be uniquely
expanded in the form
\begin{equation}
   \G_{m,n} = \sum_{a=1}^{2^d} \G^a_{m,n}\Gamma^a
   \label{eq:10}
\end{equation}
Here $m,n$ label the lattice points, $\G^a$ denotes a matrix with 
position indices, and $\Gamma^a$ is the element of the Clifford basis.
Clifford basis is built on gamma--matrices satisfying
$ \{ \gmu,\gnu \} = 2\dmunu\identity$. We will refer to the
operators $\G^a$ as Clifford components of $\G$.

It is useful to define the chirally symmetric ($\G_C$) and chirally
nonsymmetric ($\G_N$) part of arbitrary operator $\G$, using the
$\gfive$ element of the Clifford algebra, namely
\begin{equation}
   \G \;=\; {1\over 2} \gfive\, \{ \gfive,\G \} 
          + {1\over 2} \gfive\,  [ \gfive,\G  ]
               \;\equiv\; \G_N + \G_C
   \label{eq:20}
\end{equation}
This is a unique decomposition such that $\{ \G_C,\gfive \} = 0$,
and $[ \G_N,\gfive ] = 0$.

Important role in our discussion will be played by the subset
$\Gsl$ of operators that are local and respect the symmetries of the 
hypercubic lattice structure. By locality we mean at least exponential 
decay of interaction at large distances, and the symmetries of hypercubic 
lattice include translation invariance and invariance under hypercubic
rotations and reflections (hypercubic invariance). The elements 
$\G \in \Gsl$ can be equivalently represented by their Fourier image
which is diagonal
\begin{equation}
    G(p) \;\equiv\; \sum_{a=1}^{2^d} G^a(p)\Gamma^a \qquad\qquad
    G^a(p)\equiv \sum_n g_n^a e^{ip\cdot n}\qquad\qquad
    g_n^a \equiv \G^a_{0,n}
    \label{eq:30}
\end{equation}
Here the functions $G^a(p)$ of lattice momenta $p\equiv (p_1,\ldots,p_d)$ 
are complex--valued, periodic with $2\pi$ in all variables, analytic,
\footnote{As is well known from the theory of Fourier series, locality 
(as defined in Appendix~\ref{app:1}) implies that the functions $G^a(p)$, 
extended to complex momenta, are analytic in the region
$\{p \;:\; |Im(p_\mu)|< \delta, \forall \mu\,\}$ 
with some positive $\delta$. The Brillouin zone of real momenta is embedded 
in this complex region. We illustrate this in the simplest case of one
dimension in Appendix~\ref{app:3}.}      
and further restricted by hypercubic invariance. The concepts discussed 
in this paragraph are properly defined in Appendix~\ref{app:1}.

The set of lattice Dirac operators that will be considered can now be 
defined. We require locality, symmetries of the hypercubic lattice,
and correct classical continuum limit. 
\begin{definition}
   (Set $\Dset$) Let $D(p)$ be a local symmetric operator (element of $\Gsl$), 
   and let in the vicinity of $p=0$ its Clifford components $D^a(p)$ 
   satisfy
   \begin{equation}
      D^a(p) = 
        \cases{ip_\mu + O(p^2),&if $\Gamma^a  =\gmu\,$;\cr
        O(p^2),&if $\Gamma^a\ne\gmu\,,\forall\mu$\cr}
      \label{eq:40}
   \end{equation}
   Collection $\Dset\subset\Gsl$ of such elements $D(p)$ defines 
   the set of lattice Dirac operators.
\end{definition} 
Note that there are elements both with and without doublers in the set 
$\Dset$.

Finally, our analysis will focus on the property of {\it ultralocality}.
This notion reflects the fact that the operator doesn't couple 
variables beyond some fixed finite distance:
\begin{definition} (Ultralocality)
   Let ${\cal C}_N$ denotes the set of all lattice sites contained in the 
   hypercube of side $2N$, centered at $n=0$, i.e. 
   ${\cal C}_N\equiv\{n : |n_\mu|\le N,\, \mu=1,\ldots,d\}$. 
   Operator $\G$ is said to be be ultralocal if there is a positive 
   integer $N$, so that
   \begin{displaymath}
       \G^a_{m,n} \;=\; 0\,, \qquad\;          
       \forall m,n: (m-n) \not\in {\cal C}_N\,,\;\forall a
   \end{displaymath} 
\end{definition}
If an ultralocal operator $\G$ is translation invariant, then Clifford
components $G^a(p)$ of its Fourier image are functions with finite
number of Fourier terms, i.e. 
$G^a(p)\equiv \sum_{n\in{\cal C}_N} g_n^a e^{ip\cdot n}$.

\section{Formulation of the Problem}

Defining property of GW actions is the locality of chirally nonsymmetric 
part of the propagator. In other words, $\D\in \Dset$ is a GW operator 
if $(\D^{-1})_N \in \Gsl$. This is very restrictive since
the full massless propagator $\D^{-1}$ is nonlocal. GWL symmetry thus
requires that all this nonlocality is contained in the chirally symmetric
part $(\D^{-1})_C$. Actions with naive chiral symmetry can be viewed
as the limiting case in the set of GW actions as $(\D^{-1})_N$ completely 
vanishes.\footnote{
Strictly speaking, since $\D$ has zero eigenvalues (for zero 
momentum states), all the manipulations involving $\D^{-1}$ should be   
performed with small mass term added to $\D$, with massless limit taken 
at the end. This is always implicitly assumed.}

In Fourier space the GW property means that all Clifford components
of $(D^{-1}(p))_N$ are analytic functions on the whole Brillouin zone.
This is automatic for arbitrary action with naive chiral symmetry, and
can be also rather easily arranged even for ultralocal $D(p)$ with
$(D(p))_N \ne 0$,~\cite{Hor99A,Hor99C}. However, in all such known 
cases $D^{-1}(p)$ has unwanted poles at nonzero momenta, and the 
theory thus suffers from doubling. This suggests that with only finite
number of Fourier components at our disposal, it is perhaps impossible 
to add to the chirally symmetric operator $D_C$ a nonsymmetric 
part $D_N$, that would not interfere with the behavior of the 
propagator at large distances, and at the same time remove all the 
unwanted poles originating from $D_C$.

We show in the rest of this paper that this is indeed true. More
precisely, we will prove the following theorem:

\begin{theorem}
   \label{the:1}
   There is no $D(p)\in \Dset$ such that the following three
   requirements are satisfied simultaneously:   
   \begin{description}
     \item[$(\alpha)$] $D(p)$ is ultralocal. 
     \item[$(\beta)$]  $(D^{-1}(p))_N \in \Gsl$. 
     \item[$(\gamma)$] $(D^{-1}(p))_C$ has no poles except when
                       $\,p_\mu=0\!\pmod{2\pi},\, \forall\mu$. 
   \end{description} 
\end{theorem}

\section{Auxiliary Statements}

We first discuss several auxiliary statements that will 
be needed.

\subsection{Two-Dimensional Restrictions on the Brillouin Zone}

The key ingredient that leads to the successful demonstration of 
Theorem~1 rests upon the realization that one should concentrate
on the consequences of analyticity of $(D^{-1})_N$ at the origin 
of the Brillouin zone~\cite{Hor99C}. This is indeed nontrivial
because both $(D^{-1})_C$ and $(D^{-1})_N$ are functions of 
all Clifford components $D^a$, while at the same time, the former 
is singular and the latter is analytic at the origin. We will
show that to capture the required analytic structure of the 
functions involved, it is sufficient to consider two-dimensional 
restrictions on the Brillouin zone. We thus start with the 
definition of such restrictions in general.
  
\begin{definition} 
   (Restriction $\Sigma^\rho$) Let $\rho \in \{1,2,\ldots d\}$ and
   let $\pbar$ denotes the restriction of the momentum variable $p$  
   defined through 
   \begin{displaymath}
      \pbar_\mu \;=\; \cases{q_\mu,& if $\mu=1,\ldots,\rho$;\cr
                             0,& if $\mu=\rho+1,\ldots,d$\cr}
   \end{displaymath}
   Map $\Sigma^\rho$ that assigns to arbitrary function $f(p)$ of 
   $d$ real variables a function $\overline{f}(q)$ of $\rho$ real
   variables $q\equiv ( q_1,q_2,\ldots,q_\rho )$ through
   \begin{equation}
      \Sigma^\rho\Bigl[\,f(p)\,\Bigr] \;\equiv\; 
      \overline{f}(q) \;\equiv\; f(\pbar)
      \label{eq:50}
   \end{equation} 
   will be referred to as restriction $\Sigma^\rho$. 
\end{definition}

Functions with definite transformation properties under hypercubic 
group simplify or vanish when restricted through $\Sigma^\rho$. 
The elements of $\Gsl$ simplify accordingly. In particular, 
for $\Sigma^2$ we have the following.

\begin{lemma}
   \label{lem:1}
   Let $G(p)\in\Gsl$, and let $\Gr(q)$ be its restriction under 
   $\Sigma^2$ defined through 
   \begin{displaymath}
      \Gr(q) \;=\; \sum_{a=1}^{2^d} \Gr^a(q)\Gamma^a\qquad\quad  
      \Gr^a(q) \;=\; \Sigma^2\Bigl[\,G^a(p)\,\Bigr]
   \end{displaymath}
   Then $\Gr(q)$ can be written in the form
   \begin{equation}
      \Gr(q) \;=\; X(q)\identity 
                 + \sum_{\mu=1}^2 Y_\mu(q) \gmu 
                 + Z(q) \gone\gtwo
      \label{eq:60}
   \end{equation}
   where $X(q)$,$Y_\mu(q)$ and $Z(q)$ are analytic functions of two 
   real variables, periodic with $2\pi$, and satisfying
   \begin{equation}
      X(q_1,q_2) \;=\; X(-q_1,q_2) \;=\; X(q_1,-q_2) \;=\; X(q_2,q_1)
      \label{eq:70}
   \end{equation}
   \begin{equation}
      Y_1(q_1,q_2) \;=\; -Y_1(-q_1,q_2) \;=\; Y_1(q_1,-q_2) \;=\;
      Y_2(q_2,q_1)
      \label{eq:80}
   \end{equation}
   \begin{equation}
      Z(q_1,q_2) \;=\; -Z(-q_1,q_2) \;=\; -Z(q_1,-q_2) \;=\; -Z(q_2,q_1)
      \label{eq:90}
   \end{equation}      
\end{lemma}
Lemma~\ref{lem:1} reflects the naive expectation that $\Gr(q)$ formally
looks as an element of $\Gsl$ in two dimensions. The proof is given 
in Appendix~\ref{app:2}.

   The distinctive property of ultralocal elements of $\Gsl$ is that its
Clifford components are polynomials (or are related to polynomials)
in suitably chosen variables. In particular, for the restrictions 
under $\Sigma^2$ we have the following result proved in 
Appendix~\ref{app:5}.

\begin{lemma}
   \label{lem:2}
   Let $G(p)$ be an ultralocal element of $\Gsl$. Then the Clifford
   components (\ref{eq:70}-\ref{eq:90}) of its restriction $\Gr(q)$
   under $\Sigma^2$ can be written in the form
   \begin{eqnarray}
      X(q_1,q_2) &=& P_X(\cos q_1, \cos q_2)   \nonumber \\
      Y_1(q_1,q_2) &=& \sin q_1 \;P_Y(\cos q_1, \cos q_2) 
                    \;=\; Y_2(q_2,q_1) \label{eq:100} \\
      Z(q_1,q_2) &=& \sin q_1 \sin q_2 \;P_Z(\cos q_1, \cos q_2) 
      \nonumber
   \end{eqnarray}
   where $P_X$ is a symmetric polynomial in its variables, $P_Z$ an 
   antisymmetric polynomial, and $P_Y$ a polynomial. 
\end{lemma}   

\subsection{Useful Algebraic Results}

While the GW property is associated with {\it analytic} properties
of the Fourier transform of the propagator, the proof of Theorem 1 
that we give in the next section is essentially algebraic.
Lemma \ref{lem:3} below is a well-known result which will turn out 
to reflect this connection to algebra for ultralocal operators. 
Lemma~\ref{lem:4} and Corollary~\ref{cor:1} are rather interesting 
new results that we will need.  

\begin{lemma}
   \label{lem:3}
   Let $P(x,y)$, $Q(x,y)$ be two polynomials over the field of complex 
   numbers such that $Q(0,0) = 0$, and that the rational function 
   $R(x,y)\equiv P(x,y)/Q(x,y)$ is analytic in some neighborhood of the 
   origin. Then $P$ and $Q$ have common polynomial factor $F$, such
   that if we write $P=F\tilde P$, $Q=F\tilde Q$, 
   then $\tilde Q(0,0) \neq 0$. The polynomial $F$ of minimal degree
   is unique up to a constant multiplicative factor, and depends only
   on $Q$.
\end{lemma}

\begin{lemma}
  \label{lem:4}
  Let $G_1(x,y)$, $G_2(x,y)$ be polynomials over complex numbers, 
  such that $G_1(0,0) \ne 0$, $G_2(0,0) \ne 0$. Let further
  \begin{displaymath}
    B(x,y) \;\equiv\; x(1-x)\; G_1^2(x,y) + y(1-y)\; G_2^2(x,y)
  \end{displaymath}
  If $F(x,y) P(x,y) \,=\, B(x,y)$ is arbitrary polynomial 
  factorization such that
  \begin{displaymath}
     F(0,0)=0 \qquad\qquad\quad   P(0,0)\ne 0
  \end{displaymath}
  then $F(x,y)$ must have at least one additional zero on the 
  set $\cZ_0\equiv \{\, (x,y) :\, x,y \in \{0,1\}\,\}$. 
\end{lemma}

\begin{corollary}
  \label{cor:1}
  Let $G_1(x^2,y^2)$, $G_2(x^2,y^2)$ be polynomials over complex 
  numbers, such that $G_1(0,0) \ne 0$, $G_2(0,0) \ne 0$. Let further
  \begin{displaymath}
    B(x^2,y^2) \;\equiv\; 
      x^2(1-x^2)\; G_1^2(x^2,y^2) + y^2(1-y^2)\; G_2^2(x^2,y^2)
  \end{displaymath}
  If $F(x,y) P(x,y) \,=\, B(x^2,y^2)$ is arbitrary polynomial 
  factorization such that
  \begin{displaymath}
     F(0,0)=0 \qquad\qquad\quad   P(0,0)\ne 0
  \end{displaymath}
  then $F(x,y)$ must have at least two additional zeroes on the set
  $\cZ_1\equiv \{\, (x,y) :\, x,y \in \{-1,0,1\}\,\}$. 
\end{corollary}

We prove these auxiliary statements in the subsections below.\footnote{
Reader not interested in these proofs can proceed directly to the proof 
of strong non-ultralocality in Sec.~\ref{sec:6}. It should be pointed out 
however, that the proofs of Lemma~\ref{lem:4} and Corollary~\ref{cor:1}
are important for understanding the ``heart of the matter''.} 
It turns out that it is quite useful and natural for our purposes to use 
the established language of algebraic geometry. To make this paper
sufficiently self-contained, we will first summarize some basic notions 
and state some standard results that will be relevant for our purposes.

\subsubsection{Preliminaries from Algebraic Geometry}
\label{sec:sss1}

We will be repeatedly concerned here with polynomials over the field 
of complex numbers $\C$. Let us thus denote the set of such 
polynomials in indeterminates $x_1, x_2, \ldots,x_n$ 
as $\C\,[x_1, x_2,\ldots,x_n]$.\footnote{All of 
the discussion that follows can be carried over for polynomials (and 
corresponding algebraic sets) over arbitrary algebraically closed 
field. However, we will avoid such unnecessary generality here.}
The degree of arbitrary polynomial $P$ will be denoted as $\deg P$.
Also, if some expression $Q(x_1,\ldots,x_n,y_1,\ldots,y_m)$ can be
viewed as a power series in variables $x_1,\ldots,x_n$, then
$\ord_{x_1,\ldots,x_n}(Q)$ will denote the minimal degree of monomials
appearing in such series. This obviously applies also if $Q$ is
a polynomial. For example, we will work with the power series like  
$\pi(t) \equiv \sum_{i=0}^{\infty}p_it^i$, in which case $\ord_t(\pi(t))$
is the least $i$ such that $p_i \neq 0$.
\begin{definition}
   (Affine Algebraic Curve) Let $P\in \C\,[x,y]$ and $\deg P >0$. 
   The set of points in the complex affine plane 
   $\Gamma_P\,\equiv\, \{\, (x,y)\,:\, P(x,y)=0 \,\}$ defines an affine 
   algebraic curve.
   \label{def:6}
\end{definition}   
An algebraic curve is thus a geometric figure defined by the locus of 
zeroes of a polynomial. Arbitrary neighborhood of any point on the curve 
contains infinitely many other points belonging to the curve, and there
are no isolated points. If $\Gamma$ is an algebraic curve and there
exists an irreducible polynomial $P$ such that $\Gamma=\Gamma_P$, then
$\Gamma$ is called an {\it irreducible} algebraic curve. If $P(x,y)$
is reducible and $P(x,y)=\prod_{i=1}^N P_i^{n_i}(x,y)$ is a 
factorization into irreducible factors, then obviously
$\Gamma_P\,=\,\Gamma_{P_1}\union\Gamma_{P_2}\union\ldots \Gamma_{P_N}$,
and the irreducible curves $\Gamma_{P_i}$ are referred to as the
{\it components} of $\Gamma_P$. 

Introducing the geometrical viewpoint into algebraic problem frequently 
turns out to be quite useful. Note, for example, that the content of 
Lemma~\ref{lem:4} can be interpreted in the following way. To every
polynomial $B(x,y)$ constructed as prescribed we can assign the
curve $\Gamma_B$, and always $\cZ_0\subset\Gamma_B$. Similarly, to any
factor $F(x,y)$ such that $F(0,0)=0$, we can assign a curve 
$\Gamma_F\subset\Gamma_B$, running through the origin. The claim is 
that every $\Gamma_F$ has to run through one additional point 
of $\cZ_0$. To demonstrate this result, we will need some basic tools
from the theory of intersections of algebraic curves which, in turn, 
is more elegantly discussed in the {\it projective plane} rather than 
in the affine plane. 
\medskip

\noindent {\bf A. Projective Curves}
\medskip

\noindent
Two parallel lines in the affine plane do not have any points in common. 
However, it is intuitively quite acceptable to say that they meet at the
{\it ``point at infinity''}. To handle such ``intersections'' easily, it 
is useful to think about algebraic curves in the projective plane.   
\begin{definition}
   (Projective Plane) Consider the set $S={\C}^3 - \{(0,0,0)\}$ and  
   define two points ${\mathbf a},{\mathbf b}\in S$ to be equivalent 
   iff ${\mathbf a} = \lambda{\mathbf b}$ for some $\lambda\in\C$.  
   The set of equivalence classes of $S$ under this equivalence relation 
   is called the projective plane.
\label{def:7}
\end{definition}
A point in the projective plane has thus infinitely many triples of 
numbers (not simultaneously zero) associated with it. Any one of these
fully represents the point and is referred to as a triple of 
{\it homogeneous coordinates} for that point. The coordinates $\{x,y,z\}$ 
on ${\bb C}^{3}$ are thus used to define the sets of homogeneous 
coordinates for the points of projective plane. 
The subset of projective plane consisting of points represented by 
homogeneous coordinates with $z \neq 0$ is isomorphic to the affine 
plane under the map 
\begin{equation}
    \phi \;:\quad (\,x,y,z\,) \quad\longrightarrow\quad 
    (\,{x\over z},{y\over z}\,)
    \label{eq:102}
\end{equation}
Similarly, the subsets of projective plane with $x \neq 0$ and 
$y \neq 0$ are isomorphic to the affine plane under analogously defined 
maps. However, in our discussion we will use the convention fixed by 
eq.~(\ref{eq:102}).   

To understand why projective plane is useful for dealing with 
points at infinity, consider an arbitrary line in the affine plane.
This is specified by linear equation $ax + by + c = 0$, where $a,b$
are not simultaneously zero. For definiteness, assume that $b\neq 0$
so that the line can be parametrized through $x$ as 
$(\,x, -x a/b - c/b\,)$. In view of map $\phi$, this corresponds in the 
projective plane to the set of points represented by homogeneous 
coordinates $(\,x, -x a/b - c/b, 1\,)$. Concentrating on the part of 
the line when $|x|$ is very large, we can multiply by $b/x$ and use the 
homogeneous coordinates $(\,b, -a - c/x, b/x\,)$ instead. Considering 
now the limiting case $|x|\,\rightarrow\,\infty$, it is natural to 
associate the point at infinity of the affine plane corresponding 
to the above line with the point in the projective plane with homogeneous 
coordinates $(\,b,-a,0\,)$. Notice that this point is 
not part of the isomorphism realized through~(\ref{eq:102}) and there are 
no infinities associated with this point in the projective plane. Thus, 
roughly speaking, the projective plane is an affine plane together with 
its points at infinity.  
\begin{definition}
   (Projective Algebraic Curve) Let $P\in \C\,[x,y,z]$ be homogeneous 
   of nonzero degree. The set of points in the complex projective plane
   represented by homogeneous coordinates from the set  
   $\Gamma_P\,\equiv\, \{\, (x,y,z)\,:\, P(x,y,z)=0 \,\}$ 
   defines a projective algebraic curve.
   \label{def:8}
\end{definition} 
\vskip -0.23in
We deal with homogeneous polynomials in projective plane because they
simultaneously vanish for all homogeneous coordinates representing the 
same point of the projective plane. The concepts of irreducibility and 
component are defined analogously to the affine case, i.e. through 
irreducibility of the defining homogeneous polynomial.

There is a natural one to one correspondence between affine curves and 
projective curves that do not have $z=0$ (line at infinity) as a 
component. To see that, it is useful to consider the map $\Phi$ which
assigns to arbitrary $P\in \C\,[x,y]$ of degree $m$, the homogeneous
polynomial $P^*\in \C\,[x,y,z]$ of the same degree through
\begin{equation}
    \Phi \;:\quad P(x,y) \quad\longrightarrow\quad 
    P^*(x,y,z) \,\equiv\, z^m \, P(\,{x\over z}, {y\over z}\,)
    \label{eq:104}
\end{equation}
The image $P^*$ of $P$ under $\Phi$ is usually referred to as the
{\it canonical homogenization} of $P$. The following obvious theorem 
holds (see e.g. Ref.~\cite{Sei68A})
\begin{theorem}
   \label{the:2}
   $\Phi$ is a bijective map (one to one and onto) between $\C\,[x,y]$ 
   and the set $\cH \subset \C\,[x,y,z]$ of homogeneous polynomials not 
   divisible by $z$. The corresponding inverse map is given by 
   $P^*(x,y,z) \,\rightarrow\, P(x,y) \,\equiv\, P^*(x,y,1)$. Polynomial
   $P(x,y)$ is irreducible if and only if its image $P^*(x,y,z)$
   is irreducible.
\end{theorem}  
In this way bijection $\Phi$ also defines the correspondence between
affine curve $\Gamma_P$ and projective curve $\Gamma_{P^*}$, where the 
points are associated through map $\phi$ of eq.~(\ref{eq:102}). For any 
point $(x,y)$ of the affine curve $\Gamma_P$ there is a corresponding 
point on projective curve $\Gamma_{P^*}$, represented for example
by $(x,y,1)$. Conversely, for arbitrary point of the projective curve 
$\Gamma_{P^*}$ with homogeneous coordinates $(x,y,z), \; z\ne 0$, there 
is a corresponding point $(x/z,y/z)$ on $\Gamma_P$. The points on 
$\Gamma_{P^*}$ with homogeneous coordinates $(x,y,0)$ do not have the 
corresponding points in the finite affine plane and they are the
points at infinity of $\Gamma_P$. In this sense, the projective 
curve $\Gamma_{P^*}$ is frequently referred to as the 
{\it projective closure} of $\Gamma_P$. Conversely, $\Gamma_P$ is an 
{\it affine representative} of $\Gamma_{P^*}$.
\medskip
  
\noindent {\bf B. Intersections of Curves}
\medskip

\noindent 
Our discussion of curve intersections will focus on understanding 
the most basic result here, namely the Bezout's Theorem. The logic 
and formalism of our approach will be almost exclusively guided by 
its utility in the proof of Lemma~\ref{lem:4}. 

We start with the notion of {\it intersection multiplicity}, which 
is a positive integer assigned to any point where two algebraic 
curves meet. The corresponding definition is motivated by the 
concept of multiplicity for the root of a polynomial. Let us first 
consider a simple example in the affine plane. The curve 
$\Gamma_P$ cut out by the polynomial $P(x,y) = y - F(x)$ 
and the curve $\Gamma_L$ with $L(x,y)=y$ (line) meet at some point 
$(\lambda, 0)$. In this case one can easily define the intersection 
multiplicity at that point as the multiplicity of $\lambda$ as 
a root of $F$. Knowledge of intersection multiplicity assigned this 
way clearly adds a useful geometrical detail on how the two curves 
actually meet. It is instructive to think of this definition of 
intersection multiplicity also in a different way. Consider the 
parametrization $t \longrightarrow (t+\lambda, 0)$ of the line 
$\Gamma_L$. Since $t=0$ corresponds to the point $(\lambda,0)$,
such parametrization is usually referred to as parametrization
``at the point'' $(\lambda, 0)$. Upon insertion in $P(x,y)$ we get
$P(\lambda + t, 0) = -F(\lambda + t)$, and we can conclude that the
intersection multiplicity of $\Gamma_P$ and $\Gamma_L$ at the 
point $(\lambda, 0)$ is given by 
$\ord_t(F(\lambda + t)) \,=\, \ord_t(P(\lambda + t, 0))$.
We have just used an elementary fact that if $F(x)$ is a polynomial 
in ${\bb C}[x]$, then the multiplicity of $\lambda$ as a root of $F$
is given by $\ord_t(F(t+\lambda))$.  

Our aim is to generalize the above method for arbitrary curves. To do 
this, it is first necessary to put the concept of parametrization at 
a point of a curve on a firm footing. The following fundamental
statement holds~\cite{Wae39A}
\begin{theorem}
   \label{the:3}
   Let $\Gamma_P$ be arbitrary projective algebraic curve, and let
   $(x_0,y_0,z_0)$ be arbitrary point of this curve. 
   There exists a finite set of triples of functions 
   $(\mu_j(t), \nu_j(t), \xi_j(t))$, analytic in the neighborhood 
   of $t=0$, such that
   \begin{description}
     \item{(1)} $( \mu_j(0), \nu_j(0), \xi_j(0) ) \,=\, 
                 ( x_0, y_0, z_0 ) \,,\;\forall j$
     \item{(2)}  $ P(\mu_j(t), \nu_j(t), \xi_j(t)) \,=\, 0
                 \,,\;\forall j $
     \item{(3)} For every point $(x,y,z) \ne (x_0,y_0,z_0)$ of 
     $\Gamma_P$ in a suitable neighborhood of $(x_0,y_0,z_0)$, there 
     is exactly one $j$ and a unique value of $t$ such that 
     $(\mu_j(t), \nu_j(t), \xi_j(t)) = (x, y, z)$.  
   \end{description} 
\end{theorem}  
It should be emphasized that the set of triples
$(\mu_j(t), \nu_j(t), \xi_j(t))$ of Theorem~\ref{the:3} is not unique. 
In particular, we can perform an analytic change of variables 
$t=\alpha_1\tau + \alpha_2\tau^2 + \ldots\;,\, \alpha_1\neq 0$, to obtain 
an equally valid parametric representation (in $\tau$) satisfying
conditions ${\mathit (1-3)}$. For fixed $j$, the equivalence class
of such representations is usually referred to as a {\it branch}.
The point $(\mu_j(0), \nu_j(0), \xi_j(0))$ is said to be the 
{\it center} of the branch, and the branch is said to be 
{\it centered} at that point. Completely analogous considerations
hold for affine curves. As a simple example of multiple branches 
centered at a point, we can consider the curve defined by
$P(x,y) = y^2 - x^2 - x^3$, which has two branches at the origin 
given by $t \longrightarrow (\,t, t(1+t)^{1/2}\,)$ and  
$t \longrightarrow (\,t, -t(1+t)^{1/2}\,)$. 
\begin{definition}
   (Intersection Multiplicity) Let ${\mathbf a}\equiv (x_0,y_0,z_0)$
   be the intersection point of projective algebraic curves 
   $\Gamma_P$ and $\Gamma_Q$. Let further $\Gamma_P$ have $N$ branches
   $\gamma_j\equiv (\mu_j(t), \nu_j(t), \xi_j(t))$ centered 
   at ${\mathbf a}$. The intersection multiplicity of $\Gamma_P$ and 
   $\Gamma_Q$ at point ${\mathbf a}$ is defined as
   \begin{displaymath}
       i(\Gamma_P, \Gamma_Q, {\mathbf a}) \,\equiv\,
       \sum_{j=1}^N i(\gamma_j, \Gamma_Q, {\mathbf a}) \,\equiv\,
       \sum_{j=1}^N {\ord}_t\Bigl(Q(\mu_j(t), \nu_j(t), \xi_j(t))\Bigr)
   \end{displaymath}
   where $i(\gamma_j, \Gamma_Q, {\mathbf a})$ is the intersection
   multiplicity of $\gamma_j$ and $\Gamma_Q$ at point ${\mathbf a}$.
  \label{def:9}
\end{definition}
It can be shown that the above definition of 
$i(\Gamma_P, \Gamma_Q, {\mathbf a})$ is symmetric in $\Gamma_P$, 
$\Gamma_Q$, and that it is not dependent on the specific choice 
of parametrization from the corresponding equivalence classes 
defining the branches.\footnote{Intersection multiplicity at a
point is frequently introduced differently (through resultants
for example) and then the equivalence to Definition~\ref{def:9}
is shown.}

We are now finally in the position to state the fundamental theorem 
about intersections of curves in the projective plane~\cite{Sei68A}.
\begin{theorem}
  \label{the:4}
  (Bezout's Theorem) Let $\Gamma_P$ and $\Gamma_Q$ be two projective 
  algebraic curves without a common component. The total number of their 
  intersections counted with multiplicities is given by 
  $i(\Gamma_P,\Gamma_Q) \,=\, \deg P \deg Q$.
\end{theorem}

\subsubsection{Proof of Lemma~\ref{lem:3}}

\begin{proof}
Consider the factorization of $Q$ into irreducible factors, and splitting 
these factors in two groups. In particular, we label the factors that 
vanish at the origin as $F_i$, and the factors that do not vanish 
at the origin as $\tilde Q_i$. Using this, define 
$F(x,y) \equiv \prod_{i=1}^n F_i(x,y)$ and 
$\tilde Q(x,y) \equiv \prod_{i=1}^m \tilde Q_i(x,y)$. 
Note that while there is always at least one $F_i$, there doesn't 
necessarily have to be any $\tilde Q_i$. In that case we define 
$\tilde Q(x,y) \equiv 1$. We thus always have $Q = F \tilde Q$, 
with $F(0,0)=0$, and $\tilde Q(0,0) \neq 0$.

Let us look at an arbitrary but fixed $F_i$, defining an irreducible affine 
complex algebraic curve $\Gamma_{F_i}$, running through the origin.
Consider an arbitrary (complex) neighborhood of the origin, in which 
$R(x,y)$ is analytic. Any such neighborhood contains {\it infinitely}
many points of $\Gamma_{F_i}$. These points represent locations of 
potential singularities of $R(x,y)$, and thus must be canceled by zeroes 
of $P(x,y)$. However, by Bezout's theorem, any two algebraic curves 
without common component have only {\it finite} number of points in common 
so this is only possible if $\Gamma_{F_i}$ is the component of $\Gamma_P$. 
Consequently, $F_i$ must be an irreducible factor of $P$.

By the above argument, $F$ is the desired common factor of $P$ and $Q$. 
From the unique factorization theorem it follows that $F$ is the only 
factor (up to a multiplicative constant) of minimal degree. Also, its 
definition only depends on $Q$ as claimed.
\end{proof}

\subsubsection{Proof of Lemma~\ref{lem:4}}
\label{sec:sss3}

We will use the formalism described in section~\ref{sec:sss1} to view 
Lemma~\ref{lem:4} as the problem of algebraic geometry. In particular,
we will assign to the polynomials $G_1(x,y)$, $G_2(x,y)$ and $B(x,y)$
their canonical homogenizations defined by the map $\Phi$ of equation 
(\ref{eq:104}). In view of Theorem~\ref{the:2} we can then reformulate
the problem equivalently in projective terms as follows.
\begin{proposition}
  \label{pro:3}
  Let $G_1(x,y,z), G_2(x,y,z) \in \C\,[x,y,z]$ be homogeneous 
  polynomials, such that $G_1(0,0,1) \ne 0$, $G_2(0,0,1) \ne 0$. 
  Let further
  \begin{displaymath}
    B(x,y,z) \;\equiv\; x(z-x)\; G_1^2(x,y,z) + y(z-y)\; G_2^2(x,y,z)
  \end{displaymath}
  If $F(x,y,z) P(x,y,z) \,=\, B(x,y,z)$ is arbitrary polynomial 
  factorization such that
  \begin{displaymath}
     F(0,0,1)=0 \qquad\qquad\quad   P(0,0,1)\ne 0
  \end{displaymath}
  then the projective algebraic curve $\Gamma_F$ passing through
  $(0,0,1)$ must pass through at least one additional point from
  the set $\tilde{\cZ}_0\equiv \{\, (x,y,1) :\, x,y \in \{0,1\}\,\}$. 
\end{proposition}
\medskip

\begin{proof}
For convenience, let us denote $F_1(x,y,z) = x(z-x)$ and 
$F_2(x,y,z) = y(z-y)$. We will concentrate on the intersections of 
$\Gamma_F$ with $\Gamma_{F_1}$. It is sufficient to consider factors
$F$ that do not have common polynomial factor with $F_1$. Indeed, $x$
can not be a factor of $B$, and hence it can not be a factor of $F$. 
On the other hand, $z-x$ can be a factor of $F$ but if it is, than the 
claim of the proposition is obviously true.
\medskip

(I) Since $F_1$ and $F$ are mutually coprime, it follows from Bezout's
theorem that the total number of their intersections 
$i(\Gamma_{F},\Gamma_{F_1})$ counted with multiplicities is even.
\medskip 

(II) Let ${\mathbf a}$ be the intersection point of 
$\Gamma_{F}$ and $\Gamma_{F_1}$ such that 
${\mathbf a} \notin \Gamma_{F_2}$. Then the intersection multiplicity
$i(\Gamma_{F},\Gamma_{F_1},{\mathbf a})$ is even. Indeed, let
$\gamma \equiv (\mu(t), \nu(t), \xi(t))$ be arbitrary branch of
$\Gamma_F$ centered at ${\mathbf a}$. Then 
$B(\mu(t), \nu(t), \xi(t))=0$, or equivalently
\begin{equation}
   F_1 F_2 G_1^2 \,=\, - F_2^2 G_2^2 \qquad \mbox{\rm for} \qquad
   (x,y,z) \longrightarrow (\mu(t), \nu(t), \xi(t))
   \label{eq:106}
\end{equation}
Because of the squares, we can conclude that upon the substitution,
$\ord_t( F_1\,F_2 )$ is even. At the same time, since 
${\mathbf a}\notin \Gamma_{F_2}$, we have 
$F_2(\mu(0), \nu(0), \xi(0)) \ne 0$, and thus $\ord_t(F_2)=0$. 
But $\ord_t( F_1\,F_2 ) = \ord_t(F_1) + \ord_t(F_2)$,
and so $\ord_t(F_1) = i(\gamma,\Gamma_{F_1},{\mathbf a})$ 
is a positive even integer. Since this is true for all branches 
$\gamma$ of $\Gamma_F$ centered at ${\mathbf a}$, we can conclude 
that $i(\Gamma_{F},\Gamma_{F_1},{\mathbf a})$ is even as claimed.
\medskip

(III) If $\bfa=(0,0,1)$, then 
$i(\Gamma_{F},\Gamma_{F_1},\bfa)=1$. Indeed, $\Gamma_{F_1}$
has a single branch centered at $\bfa$, namely the line that 
can be parametrized as $\lambda=(0,t,1)$. Since 
$P(0,0,1) \ne 0$ and $G_2(0,0,1) \neq 0$ we have
\begin{displaymath}
    i(\Gamma_{F},\Gamma_{F_1},\bfa_1) \,=\, 
    {\ord}_t\Bigl(F(0,t,1)\Bigr) \,=\, 
    {\ord}_t\Bigl(F(0,t,1)P(0,t,1)\Bigr) \,=\,  
    {\ord}_t\Bigl(t(1-t)G_2^2(0,t,1)\Bigr) \,=\,1
\end{displaymath}

For (I-III) to be satisfied simultaneously, $\Gamma_F$ and 
$\Gamma_{F_1}$ must necessarily intersect at one additional point
from the set of points where $\Gamma_{F_1}$ and $\Gamma_{F_2}$
meet. However, this set is precisely represented by the
set of homogeneous coordinates from $\tilde{\cZ}_0$. Consequently, 
$\Gamma_F$ has to pass through one additional point
from $\tilde{\cZ}_0$ as claimed. 
\end{proof}

\subsubsection{Proof of Corollary~\ref{cor:1}}

\begin{proof} Let us consider an arbitrary but fixed polynomial 
$B(x^2,y^2)$ of the prescribed form, and the set of all 
factorizations $B(x^2,y^2) = F(x,y) P(x,y)$ with required properties.
It will be sufficient to prove the statement for the case when 
$F(x,y)$ does not contain any nontrivial polynomial factor with nonzero 
constant term. Indeed, if $F(x,y)$ has such a factor $P_1(x,y)$, 
then redefining $F(x,y) \rightarrow F(x,y)/P_1(x,y)$, 
$P(x,y) \rightarrow P(x,y) P_1(x,y)$ can only make the set
of zeroes for new $F(x,y)$ smaller. As a consequence of the 
unique factorization theorem, the factors $F$, $P$ are unique up to 
a trivial rescaling by complex number if this additional condition 
is imposed.

Assuming the above, consider the transformation $x \rightarrow -x$, 
which does not change $B(x^2,y^2)$. Since all possible nontrivial 
irreducible factors $F_i(x,y)$ of $F(x,y)$ satisfy $F_i(0,0)=0$, 
while all irreducible factors $P_i(x,y)$ of $P(x,y)$ have 
$P_i(0,0) \ne 0$, it follows from the unique factorization theorem
that $F(x,y) \rightarrow F(x,y)$ and $P(x,y) \rightarrow P(x,y)$
under the above reflection. The same is true under 
$y \rightarrow -y$ and, as a result, $F=F(x^2,y^2)$, $P=P(x^2,y^2)$.
This implies that we can change the variables $x^2 \rightarrow x$,
$y^2 \rightarrow y$ in the whole problem. In new variables, we 
have  
\begin{displaymath}
    B(x,y) \;\equiv\; x(1-x)\,G_1^2(x,y) + y(1-y)\,G_2^2(x,y)  
          \;=\; F(x,y)\,P(x,y)
\end{displaymath}
with properties at the origin preserved. However, according to
Lemma~\ref{lem:4}, $F(x,y)$ must have an additional zero on the
set $\cZ_0\equiv \{\, (x,y) :\, x,y \in \{0,1\}\,\}$. 
Consequently, $F(x^2,y^2)$ must vanish for at least two additional
points from the set 
$\cZ_1\equiv \{\, (x,y) :\, x,y \in \{-1,0,1\}\,\}$ which completes 
the proof.
\end{proof}

\section{Proof of Strong Non-Ultralocality (Theorem~\ref{the:1})}
\label{sec:6}

We now use the formalism of Sec.~2 and auxiliary statements of Sec.~4
to prove {\it ``strong non-ultralocality''} of GW actions (Theorem 1).
\medskip

\begin{proof}
We will proceed by contradiction, and thus first assume that there 
actually exists an element $D(p) \in \Dset$ such that the requirements
$(\alpha-\gamma)$ are satisfied. To this $D(p)$ we will assign its 
restriction $\Dr(q)$ under $\Sigma^2$ which, according to 
Lemma~\ref{lem:1}, has the form $(\, q\equiv (q_1,q_2)\,)$
\begin{equation}
   \Dr(q) \;=\; \oA(q)\,\identity \,+\, 
          i\sum_{\mu=1}^2 \oB_\mu(q)\,\gmu \,+\,
          \oC(q)\, \gone\gtwo
   \label{eq:120}
\end{equation}
with the corresponding propagator being
\begin{equation}
   \label{eq:130}
   \Dr^{-1} \;=\; 
           {{\;\oA\;} \over {\;\oQ\;}} \, \identity \,-\, 
           i\sum_{\mu=1}^2 {{\,\oB_\mu\,}\over {\;\oQ\;}} \,\gmu \,-\,
           {{\;\oC\;}\over {\;\oQ\;}} \, \gone\gtwo    \qquad\quad
   \oQ \;\equiv\; \oA^2 + \oB_\mu\oB_\mu + \oC^2        
\end{equation}
Note that according to conditions $(\beta)$ and $(\gamma)$, no Clifford
component of $\Dr^{-1}$ has a pole away from the origin of the Brillouin
zone. Consequently, the function
\begin{equation}
     \label{eq:134}
     \Bigl({{\;\oA\;} \over {\;\oQ\;}}\Bigr)^2 \,+\,
     \Bigl({{\,\oB_\mu\,}\over {\;\oQ\;}}\Bigr) 
     \Bigl({{\,\oB_\mu\,}\over {\;\oQ\;}}\Bigr) \,+\,
     \Bigl({{\;\oC\;} \over {\;\oQ\;}}\Bigr)^2 \;=\; 
     {{\;1\;} \over {\;\oQ\;}}
\end{equation}
has also no poles away from the origin. 
\medskip

Since $\Dr(q)$ is ultralocal, we can use Lemma~\ref{lem:2}, to conclude
that its Clifford components have the following structure
\begin{eqnarray}
     \label{eq:140}
     \oA(q_1,q_2) &=& P_{\oA}(\cos q_1, \cos q_2) \nonumber \\
     \oB_1(q_1,q_2) &=& \sin q_1\, P_{\oB}(\cos q_1, \cos q_2) 
         \;=\; \oB_2(q_2, q_1)  \\ 
     \oC(q_1,q_2) &=& 
     \sin q_1 \sin q_2\, P_{\oC}(\cos q_1, \cos q_2) \nonumber
\end{eqnarray}
where $P_{\oA}$ is a symmetric polynomial, $P_{\oC}$ an antisymmetric
polynomial, and $P_{\oB}$ a polynomial without definite symmetry
properties. The classical continuum limit (\ref{eq:40}) requires
that $P_{\oA}(1,1)=0$, $P_{\oB}(1,1)=1$, and also $P_{\oC}(1,1)=0$ by 
virtue of its antisymmetry.
\medskip

The GW property $(\beta)$ demands that the Clifford components of 
chirally nonsymmetric part of the propagator ($\oA/\oQ$ and $\oC/\oQ$) 
are {\it analytic} in some complex region containing the (real) 
Brillouin zone. We will concentrate on the consequences of analyticity 
in the vicinity of the origin. To this end, it is convenient to 
introduce a change of variables such as 
\begin{equation}
   x \;=\; \sin\,{q_1 \over 2} \qquad\qquad
   y \;=\; \sin\,{q_2 \over 2}
\end{equation}
which is invertible in the vicinity of the origin, maps the real region 
$[-\pi,\pi]\times[-\pi,\pi]$ onto the square $[-1,1]\times[-1,1]$, 
and preserves analyticity in new variables except possibly on the 
boundary. Using (\ref{eq:140}), we have 
\begin{eqnarray}
     \label{eq:160}
     \oA(q_1,q_2) &=& A(x^2, y^2) \nonumber \\
     \oB_1(q_1,q_2) &=& x\,\sqrt{1-x^2}\, G(x^2, y^2) 
         \;=\; \oB_2(q_2, q_1)  \\ 
     \oC(q_1,q_2) &=& 
     \sqrt{\mathstrut 1-x^2} \, \sqrt{\mathstrut 1-y^2} \,xy\, 
     C(x^2, y^2) \nonumber
\end{eqnarray}
where $A$ is a symmetric polynomial such that $A(0,0)=0$,  
$C$ is an antisymmetric polynomial, and $G$ is a 
polynomial such that $G(0,0)=2$. It is also convenient
to introduce symmetric polynomials
\begin{equation}
      \label{eq:170}
      B(x^2,y^2) \;\equiv\; 
      x^2(1-x^2)\, G^2(x^2,y^2) \,+\, y^2(1-y^2)\, G^2(y^2,x^2)
\end{equation}

\vskip -0.35cm
\begin{equation}
      \label{eq:180}
      Q(x^2,y^2) \;\equiv\;  
      A^2(x^2,y^2) \,+\, B(x^2,y^2) \,+\, 
      x^2 y^2 (1-x^2) (1-y^2)\,C^2(x^2,y^2) 
\end{equation}
corresponding to $\oB_\mu \oB_\mu$, and $\oQ$ respectively. With 
this notation, the GW property $(\beta)$ implies that the functions
\begin{displaymath}
   R_1(x^2,y^2) \,\equiv\, {A(x^2,y^2) \over Q(x^2,y^2)} \qquad\quad
   R_2(x,y) \,\equiv\, 
      \sqrt{\mathstrut 1-x^2} \; \sqrt{\mathstrut 1-y^2} \; xy \;
      {{C(x^2,y^2)} \over {Q(x^2,y^2)}}
\end{displaymath}
are analytic in the vicinity of the origin.
\medskip

Let us first consider $R_1$ which is a {\it rational function} whose 
denominator vanishes at the origin. Consequently, according to 
Lemma~\ref{lem:3}, the polynomials $A$ and $Q$ have a common polynomial
factor $F(x,y)$ with zero at the origin, so that the possible
singularity is removed. Let us fix $F$ to be of minimal degree, and thus 
unique up to a constant multiplicative factor. Next, consider $R_2$. 
Since $\sqrt{1-x^2}$ and $\sqrt{1-y^2}$ are analytic and nonzero near the 
origin, the rational function $xyC/Q$ is also analytic and, moreover, 
the same denominator $Q$ is involved as in $R_1$. Consequently, 
the polynomial $F$ divides $xy\,C$. In summary, we thus have that $F$ 
divides $A$,$\;xy C$ and $Q$. Hence, from the form 
(\ref{eq:180}) of $Q$ it follows that $F$ also divides $B$.
\medskip

According to Corollary~\ref{cor:1}, an intriguing property of any 
polynomial $B(x^2,y^2)$ of the form (\ref{eq:170}) is that if
it is divisible by polynomial $F(x,y)$ vanishing at the origin, 
then $F$ has at least two additional zeroes on the set of points 
$\cZ_1 = \{\, (x,y) :\, x,y \in \{-1,0,1\}\, \}$. Since $F$
divides $Q$, it follows that $Q$ also has these zeroes on $\cZ_1$. 
Returning back to momentum variables, we thus came to the conclusion 
that in addition to the origin of the Brillouin zone, the function 
$\oQ(q)$ of equation (\ref{eq:130}) has at least two additional 
zeroes on the set 
$\tilde\cZ_1 = \{\, (q_1,q_2) :\, q_1,q_2 \in \{-\pi,0,\pi\}\, \}$.
However, this contradicts the conclusion of (\ref{eq:134}) that 
$1/\oQ$ has no poles away from the origin of the Brillouin
zone. That completes the proof.
\end{proof}  

We have thus demonstrated that every ultralocal GW action with
hypercubic symmetries has a doubler. The argument given above can be
refined using the statement of Proposition~\ref{pro:4}, formulated
and proved in Appendix~\ref{app:4}. This implies that there is always 
a doubler at the corner $(\pi,\pi)$ of the restricted Brillouin zone.

\section{Conclusions}

As a result of several breakthroughs in the last decade, lattice 
field theory appears to have finally reached the stage when it
can deal with all the important symmetries relevant in particle 
physics. In particular, it is now conceptually quite clear how to 
formulate at least vectorlike lattice models simultaneously possessing 
the lattice counterparts of gauge symmetry (Wilson gauge symmetry), 
Poincar\'e symmetry (symmetries of hypercubic lattice) and chiral 
symmetry (GWL symmetry), so that the continuum limit with desired 
field-theoretic properties can be taken comfortably.
The path of developments leading to this point essentially coincides
with the attempts to better understand lattice fermionic actions.
Fig.~1 is a graphical representation of the basic knowledge that 
we now have~\cite{Hor99C}, including the result on strong 
non-ultralocality demonstrated here.

The base set $A$ in Fig.~1 represents all actions quadratic
in fermionic variables, that have ``easy symmetries'' 
(gauge and hypercubic symmetries), and proper classical 
continuum limit. In other words, it is the set of fermionic actions 
described by lattice Dirac kernels that are covariant under symmetry
transformations of hypercubic lattice, gauge covariant, and with 
classical limit corresponding to continuum Dirac operator. 
The base set $A$ is split into three parts 
$A = A^U \union A^L \union A^N$, representing the actions that are
ultralocal, local but not ultralocal, and nonlocal respectively. 
Highlighted is also the subset of actions without doubling of species. 
Obviously, of prime interest for physics applications is the set 
$A^{ND} \intersection (A^U \union A^L)$.

As a consequence of Nielsen-Ninomiya theorem, the relation 
of the subset $A^C$ of actions with naive chiral symmetry to the
above defined sets is as indicated. While it is quite easy to construct
nonlocal elements of $A^C$ without doublers, there is no intersection of
$A^C$ with $A^{ND}$ on the local part of the diagram. The suggestion of
Ginsparg and Wilson was that there might be a larger set of actions 
contained in $A$, respecting the chiral dynamics properly. Recent 
results in the field confirmed this idea and, more importantly, lead to 
the conclusion that fermion doubling is not a definite property of 
local GW actions, i.e. 
$A^L \intersection A^{GW} \intersection A^{ND} \neq \emptyset$ (see the
filled area in Fig.~1).     
However, as a consequence of strong non-ultralocality, doubling
is a definite property of ultralocal GW actions and
$A^U \intersection A^{GW} \intersection A^{ND} = \emptyset$. 
Unless the set of actions with acceptable chiral dynamics can be further
enlarged, non-ultralocality can thus be viewed as a necessary condition 
to reconcile chiral dynamics with proper anomaly structure in lattice 
gauge theories respecting the symmetries of the hypercubic lattice.

\begin{figure}
\epsfxsize=13.0cm
\centerline{\epsffile{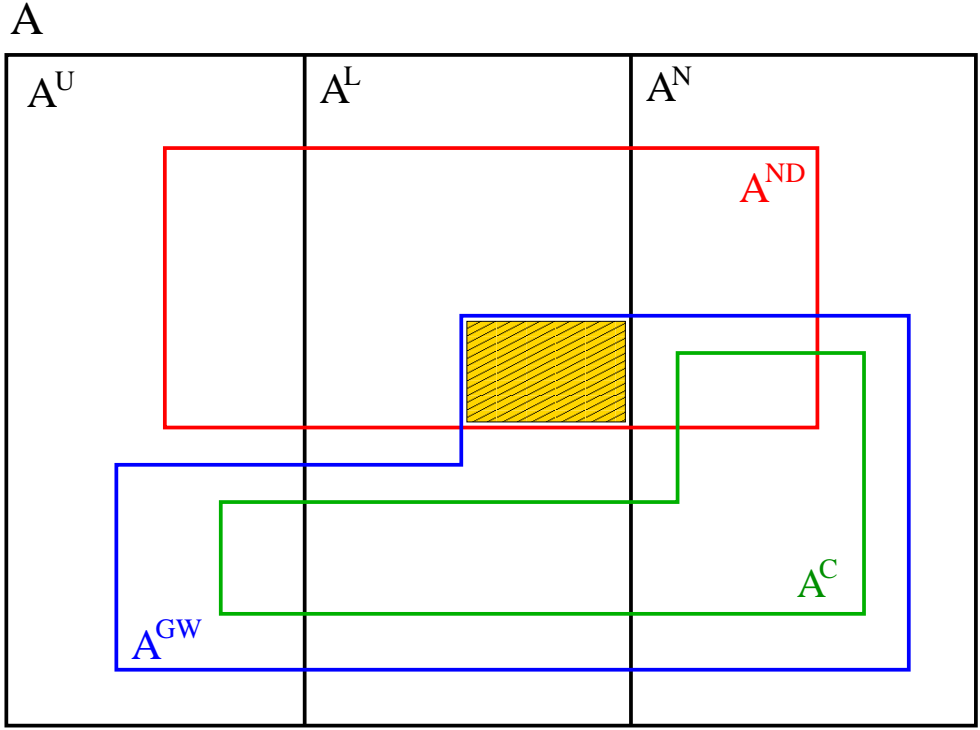}}
\caption{Lattice fermionic actions. Indicated are the sets with following
         properties: $A$ - hypercubic symmetries, gauge invariance,
         relativistic classical continuum limit; $A^U$ - ultralocal,
         $A^L$ - local but not ultralocal, $A^N$ - nonlocal, 
         $A^{ND}$ - no doublers, $A^C$ - naive chiral symmetry, 
         $A^{GW}$ - GWL symmetry.}
\vspace{-0.15cm}
\end{figure}

We would also like to stress that weak non-ultralocality and strong
non-ultralocality are two independent statements in the sense that
one does not follow from the other. In this respect it should be
pointed out that while non-ultralocality of GWL transformations applies 
regardless of doubling and does not apply for actions with naive chiral 
symmetry, non-ultralocality of GW actions is only necessary for 
doubler-free theories and actions with naive chiral symmetry represent 
no exception for this case.

\bigskip\medskip
\noindent
{\bf Acknowledgements:} I.~H. and C.~T.~B. are grateful to William Fulton
for his interest, help and support regarding Lemma~\ref{lem:4}. 
R.~M. would like to acknowledge useful input and conversations
with Dan Burghelea. I.~H. benefited from communications with 
Pavel B\'ona, Martin Niepel, Franti\v sek Marko, Arthur Mattuck and Andrei 
Rapinchuk, as well as from many pleasant discussions with Ziad Maassarani 
and Hank Thacker. We also thank Ziad for reading the manuscript.

\begin{appendix}

\section{Local Symmetric Operators}
\label{app:1}

The elements of the set $\Gsl$ satisfy three requirements defined below:

\begin{definition} 
(Locality) Operator $\G$ is said to be local if there are positive real 
constants $c$, $\delta$ such that all its Clifford components $\G^a$
satisfy
\begin{displaymath}
   |\G^a_{m,n}| \;<\; c\, e^{-\delta |m-n|}\qquad\quad
   \forall m,n
\end{displaymath}
Here $|m-n|$ denotes the Euclidean norm of $m-n$.
\end{definition} 

\begin{definition} 
(Translation Invariance) Operator $\G$ is said to be translationally 
invariant if all its Clifford components $\G^a$ satisfy 
\begin{displaymath}
    \G_{m,n}^a = \G_{0,n-m}^a\equiv g_{n-m}^a \qquad\quad
    \forall m,n
    \label{eq:500}
\end{displaymath}
\end{definition}

\begin{definition} 
(Hypercubic Invariance) Let $\Hcal$ be an element of the hypercubic group
in defining representation and $H$ the corresponding element of the
representation induced on hypercubic group by spinorial representation of 
$O(d)$. Operator $\G$ is said to have hypercubic invariance if for 
arbitrary $\Hcal$, $m$, $n$ we have
\begin{displaymath}
   \G_{n,m} \;=\; H^{-1} \G_{\Hcal n, \Hcal m}\, H
\end{displaymath}
\end{definition}

In the Fourier space, we can directly define:

\begin{definition} (Set $\Gsl$)  
  Let $G^a(p),\; a=1,2,\ldots 2^d$, are the complex valued functions of real 
  variables $p_\mu$, and let $G(p)$ be the corresponding matrix function 
  constructed as in Eq.~{\rm (\ref{eq:30})}. We say that $G(p)$ belongs 
  to the set $\Gsl$ if:
     \begin{description}
       \item[$(\alpha)$] Every $G^a(p)$ is an analytic function with 
       period $2\pi$ in all $p_\mu$.
       \item[$(\beta)$] For arbitrary hypercubic transformation $\Hcal$
       it is true identically that
       \begin{equation}
          G(p) \;=\; \sum_{a=1}^{2^d} G^a(p)\Gamma^a  
          \;=\; \sum_{a=1}^{2^d} G^a(\Hcal p) H^{-1}\Gamma^a H 
          \label{eq:510}
       \end{equation}
     \end{description}
\end{definition}

Arbitrary hypercubic transformation $\Hcal$ can be decomposed into 
products of reflections of single axis (${\cal R}_\mu$) and exchanges 
of two different axis (${\cal X}_{\mu\nu}$). Transformation properties of 
the elements of the Clifford basis are determined by the fact that $\gmu$
transforms as a vector. In particular
\begin{displaymath}
  R_\nu^{-1}\gmu R_\nu \;=\; \cases{-\gmu,& if $\mu=\nu$;\cr
                                     \gmu,& if $\mu\ne\nu$\cr}
\end{displaymath}
and
\begin{displaymath}
  X_{\rho\sigma}^{-1}\gmu X_{\rho\sigma} \;=\;
                 \cases{\gamma_\sigma,& if $\mu=\rho$;\cr
                        \gamma_\rho,&   if $\mu=\sigma$;\cr
                        \gmu,&          otherwise\cr}
\end{displaymath}
where $R_\mu, X_{\mu\nu}$ are the spinorial representations of 
${\cal R}_\mu, {\cal X}_{\mu\nu}$. The elements of the Clifford 
basis naturally split into groups with definite transformation 
properties and the hypercubic symmetry thus translates into definite 
algebraic requirements on functions $G^a(p)$.

\section{Locality and Analyticity}
\label{app:3}

\begin{proposition}
\label{pro:2}
  Let $G(p)$ be a complex function of single variable periodic with 
  $2\pi$, and forming a Fourier pair with the sequence of complex numbers
  $\{ g_n\,;\, n=0,\pm 1,\pm 2 \ldots \}$, i.e.
  \begin{displaymath}
      G(p) \;=\; \sum_{n=-\infty}^{\infty} g_n e^{i\,n\,p} \;\equiv\;
      g_0 + g_1 e^{i\,p} + g_{-1} e^{-i\,p} + 
      g_2 e^{i\,2\,p} + g_{-2} e^{-i\,2\,p} + \ldots
  \end{displaymath}
  and
  \begin{displaymath}
      g_n \;=\; {1 \over {2\pi}} \int_{-\pi}^{\pi} dp\, e^{-i\,n\,p}\;G(p)
  \end{displaymath}
  Then the following two statements are equivalent: 
  \medskip
 
  \noindent 
  (a) There exist $c>0$,$\,\delta>0$ such that 
      $\, \mid g_n \mid \;<\; c\,e^{-\delta\, |n|},\;\forall\, n$. 
  \smallskip 

  \noindent
  (b) There exists $\Delta > 0$, such that $G(p)$ is analytic in the
      complex strip $\mid Im(p)\mid < \Delta$.
\end{proposition}
\medskip
\begin{proof}
We first show that (a) implies (b). 
Consider the change of variables $z = e^{ip}$. Then $G=g(e^{ip})$ where 
the function $g$ is defined by
   \begin{displaymath}
   g(z) = g_0 + g_1 z + g_{-1} z^{-1}  + g_2 z^2 + g_{-2} z^{-2} + \cdots
   \end{displaymath} 
We split this up as $g(z) = h(z) + k(z)$ where
   \begin{displaymath}
    h(z) = g_0 + g_1 z + g_2 z^2 + \ldots  \qquad\quad
    k(z) = g_{-1}z^{-1} + g_{-2}z^{-2} + \ldots 
   \end{displaymath}
Condition (a) simply says that $|g_n| < c\beta^{-|n|}$ where
$\beta > 1$, implying that $h(z)$ has a radius of convergence 
strictly greater than $1$. Similarly, if we further change the variable 
$z=1/t$ in $k(z)$, so that $l(t) = k(1/t)$, we can infer that $l(t)$ too 
has a radius of convergence strictly greater than $1$. The above 
conclusions prove that $g(z)$ is analytic in an annulus containing the 
unit circle of the complex plane. Consequently, $G(p)=g(e^{ip})$ is 
a composition of two analytic maps, implying analyticity in the complex 
strip as desired in (b).

Next we show that (b) implies (a). Consider arbitrary 
$0 < \rho < \Delta$. We will integrate in the complex plane along the 
rectangular contour $\cC$ with the following line segments: 
$\cC_1$ from point $(-\pi,0)$ to point $(\pi,0)$, $\,\cC_2$ from  
$(\pi,0)$ to $(\pi,-\rho)$, $\,\cC_3$ from $(\pi,-\rho)$ to 
$(-\pi,-\rho)$, and $\,\cC_4$ from $(-\pi,-\rho)$ to $(-\pi,0)$.

Let us assume that $n\geq 0$. According to Cauchy theorem, we have
   \begin{displaymath}
     \int_{\cC} dp\, e^{-i\,n\,p}\;G(p) \;=\; 0 \;=\;
     \int_{\cC_1} \,+\, \int_{\cC_2} \,+\, \int_{\cC_3} \,+\, \int_{\cC_4}
   \end{displaymath}
Since $G(p)$ is periodic with $2\pi$, the contributions from the 
integrals along $\cC_2$ and $\cC_4$ will cancel each other and we
thus obtain 
   \begin{eqnarray*}
     \int_{\cC_1} dp\, e^{-i\,n\,p}\;G(p) & = &
     \int_{-\pi}^{\pi} dt \, e^{-i\,n\,t}\;G(t) \;=\;
     -\int_{\cC_3} dp\, e^{-i\,n\,p}\;G(p) \;=\;       \\      
     & = & \int_{-\pi}^{\pi} dt \, e^{-i\,n\,(t-i\rho)}\;G(t-i\rho) \;=\;
     e^{-n\,\rho}\, \int_{-\pi}^{\pi} dt \, e^{-i\,n\,t}\;G(t-i\rho)
   \end{eqnarray*}  
Finally, $G(p)$ is bounded on the path of the last integral due to analyticity, 
and we thus have
   \begin{displaymath}
      \mid g_n \mid \quad = \quad 
      \mid {1 \over {2\pi}} \int_{-\pi}^{\pi} dt\, e^{-i\,n\,t}\;G(t) \mid 
      \quad \leq \quad 
      e^{-n\,\rho} \max_{t\in [-\pi,\pi]} \mid G(t-i\rho) \mid
   \end{displaymath}

Similarly if $n<0$, then we will use the rectangular integration contour
in the upper half of the complex plane, yielding an analogous bound.
Together, this then implies (a) as claimed.
\end{proof}

\section{Proof of Lemma~\ref{lem:1}}
\label{app:2}

\begin{proof}
Let us denote the set of indices $u^\rho\equiv \{1,2\ldots \rho\}$ for
arbitrary positive integer $\rho$. Clifford basis can be subdivided into 
non--intersecting subsets $\Gamma=\union_j \Gamma_{(j)}$, 
where $\Gamma_{(j)}$, $j=0,1,\ldots, d$, contains the elements that can 
be written as the product of $j$ gamma-matrices. In particular, 
$\Gamma_{(0)}=\{\,\identity\,\}$, 
$\Gamma_{(1)}=\{\,\gmu, \;\mu\in u^d\,\}$, 
and so on. With the appropriate convention on ordering of 
gamma--matrices in the definition of $\Gamma^a$, we can then rewrite the
Clifford decomposition of $G(p)$ in the form
\begin{equation}
    G(p) \;=\; \sum_{j=0}^d \sum_{\mu_1,\mu_2\ldots\mu_j\atop
                                  \mu_1<\mu_2\ldots<\mu_j}
               F_{\mu_1,\mu_2\ldots\mu_j}(p)\, 
               \gamma_{\mu_1}\gamma_{\mu_2}\ldots\gamma_{\mu_j}\,,
               \qquad\quad \mu_i\in u^d
    \label{eq:520}
\end{equation}

We will concentrate on contributions to $\Gr(q)$ originating from subsets 
$\Gamma_{(j)}$, where $j\ge 2$. Consider a single term in decomposition 
(\ref{eq:520}) from this group, specified by the set of indices 
$v\equiv\{\mu_1,\mu_2\ldots\mu_j\} \ne u^2$. Then there exists element 
$\mu\in v$, such that $\mu\notin u^2$. Under reflection ${\cal R}_\mu$
through the corresponding axis, we have 
$
R_\mu^{-1} \,\gamma_{\mu_1}\gamma_{\mu_2}\ldots\gamma_{\mu_j}\, R_\mu
\,=\, -\gamma_{\mu_1}\gamma_{\mu_2}\ldots\gamma_{\mu_j}\,.
$
Hypercubic symmetry of $G(p)$ then requires that
$
 F_{\mu_1,\mu_2\ldots\mu_j}({\cal R}_\mu p) =
 -F_{\mu_1,\mu_2\ldots\mu_j}(p)\,.
$
However, since $\mu\notin u^2$, the restricted variable $\pbar$ under 
$\Sigma^2$ satisfies ${\cal R}_\mu \pbar = \pbar$, and hence 
\begin{displaymath}
   \Fr_{\mu_1,\mu_2\ldots\mu_j}(q) \;\equiv\;
   F_{\mu_1,\mu_2\ldots\mu_j}(\pbar) \;=\; -  
   F_{\mu_1,\mu_2\ldots\mu_j}(\pbar) \;=\; 0
\end{displaymath}
Consequently, the only Clifford element contributing from this group
is $\gone\gtwo$ (when $v=u^2$), and the form~(\ref{eq:60}) follows.

Analyticity, periodicity and relations~(\ref{eq:70}-\ref{eq:90}) for 
functions $X(q),Y_\mu(q),Z(q)$ follow from corresponding properties
of unrestricted operator. 
\end{proof}

\vfill\eject

\section{Proof of Lemma~\ref{lem:2}}
\label{app:5}

\begin{proof}
   Since $G(p)$ is ultralocal, the Clifford components of $\Gr(q)$
   have finite number of Fourier terms. Then there exists a 
   non-negative integer $N$, such that when grouping together
   the Fourier terms related by reflection properties in 
   (\ref{eq:70}-\ref{eq:90}) of Lemma~\ref{lem:1}, the Fourier 
   expansions can be written in the form
   \begin{eqnarray*}
      X(q_1,q_2)   &=& \sum_{n_1=0}^N \sum_{n_2=0}^N
                       x_{n_1 n_2} \cos n_1q_1 \cos n_2q_2
                       \nonumber \\ 
      Y_1(q_1,q_2) &=& \sum_{n_1=1}^N \sum_{n_2=0}^N
                       y_{n_1 n_2} \sin n_1q_1 \cos n_2q_2 
                       \;=\; Y_2(q_2,q_1)\\
      Z(q_1,q_2)   &=& \sum_{n_1=1}^N \sum_{n_2=1}^N
                       z_{n_1 n_2} \sin n_1q_1 \sin n_2q_2
                       \nonumber
   \end{eqnarray*}
   Furthermore, the exchange properties in (\ref{eq:70}) and
   (\ref{eq:90}) imply that $x_{n_1,n_2} = x_{n_2,n_1}$, while
   $z_{n_1,n_2} = -z_{n_2,n_1}$. Using the formulas for trigonometric 
   functions of multiple arguments, namely
   \begin{eqnarray*}
     \sin n x \;=\; \sin x \,\Biggl[\, 
       2^{n-1}\cos^{n-1}x &-&
       {{n-2}\choose 1} 2^{n-3} \cos^{n-3}x \;+\;
       {{n-3}\choose 2} 2^{n-5} \cos^{n-5}x \\ 
       &-& {{n-4}\choose 3} 2^{n-7} \cos^{n-7}x \;+\; 
       \ldots \;\Biggr]
   \end{eqnarray*}        
   and
   \begin{eqnarray*}
     \cos n x \;=\; 
       2^{n-1}\cos^n x &-&
       \frac{n}{1}\,2^{n-3} \cos^{n-2}x \;+\; 
       \frac{n}{2}\,{{n-3}\choose 1} 2^{n-5} \cos^{n-4}x \\        
       &-& \frac{n}{3}\,{{n-4}\choose 2} 2^{n-7} \cos^{n-6}x 
       \;+\; \ldots
   \end{eqnarray*}        
   the forms (\ref{eq:100}) directly follow, together with the 
   exchange symmetry properties.
\end{proof}

\vfill\eject

\section{Refined Position of the Doubler}
\label{app:4}

\begin{proposition}
  \label{pro:4}
  Let $G(x,y,z) \in \C\,[x,y,z]$ be homogeneous polynomial, such 
  that $G(0,0,1) \ne 0$. Let further
  \begin{displaymath}
    B(x,y,z) \;\equiv\; x(z-x)\; G^2(x,y,z) + y(z-y)\; G^2(y,x,z)
  \end{displaymath}
  If $F(x,y,z) P(x,y,z) \,=\, B(x,y,z)$ is arbitrary polynomial 
  factorization such that
  \begin{displaymath}
     F(0,0,1)=0 \qquad\qquad\quad   P(0,0,1)\ne 0
  \end{displaymath}
  then the projective algebraic curve $\Gamma_F$ passing through
  $(0,0,1)$ must also pass through $(1,1,1)$. 
\end{proposition}

\begin{proof} 
   We are dealing with the special case of Proposition~\ref{pro:3}, 
   given by $G_1(x,y)=G_2(y,x)=G(x,y)$, and all the arguments of the 
   the proof in section~\ref{sec:sss3} are valid here as well. Using
   the notation defined there and setting 
   $\bfa_1=(0,0,1)\in \tilde{\cZ}_0$ we have in particular 
   \begin{equation}
        \sum_{\bfa \in \tilde{\cZ}_0} i(\Gamma_{F},\Gamma_{F_1},\bfa)
        \;=\; 0 \pmod{2}  \qquad\qquad\qquad
        i(\Gamma_{F},\Gamma_{F_1},\bfa_1) = 1
        \label{eq:540} 
   \end{equation}
   We will show below that for $\bfa_2=(1,0,1),\,\bfa_3=(0,1,1)$
   it is now true in addition that
   \begin{equation} 
      i(\Gamma_{F},\Gamma_{F_1},\bfa_2) \,+\, 
      i(\Gamma_{F},\Gamma_{F_1},\bfa_3) \,=\, 0 \pmod{2}
      \label{eq:550}
   \end{equation}
   However, denoting $\bfa_4=(1,1,1)$ we have 
   $\tilde{\cZ}_0=\{\bfa_1,\bfa_2,\bfa_3,\bfa_4\}$ and so 
   (\ref{eq:540}), (\ref{eq:550}) imply that 
   $i(\Gamma_{F},\Gamma_{F_1},\bfa_4)=1 \pmod{2}$. That  
   proves the claim of the proposition.  

   To show (\ref{eq:550}), it is sufficient to concentrate on irreducible
   $\Gamma_F$. This is because such $\Gamma_F$ is unique and all the 
   reducible ones will thus contain it. The uniqueness of irreducible
   $\Gamma_F$ follows from the fact that if there were at least two, 
   then $\ord_{x,y}(B(x,y,z))$ would also have to be at least two, 
   which is not the case. This also means that the corresponding 
   $F(x,y,z)$ is symmetric in $x,y$. Indeed, if it were not, then the 
   polynomial $F(y,x,z)$ would define another component of $\Gamma_B$ 
   running through the point $(0,0,1)$ thus contradicting the uniqueness.

   Assuming the above, consider arbitrary branch 
   $\gamma=(\mu(t),\nu(t),1)$ of $\Gamma_F$, centered at $\bfa_2$. 
   Due to symmetry in $x,y$ there is a corresponding branch 
   $\gamma' = (\nu(t),\mu(t),1)$ centered at $\bfa_3$. Using the fact 
   that $\ord_t(\mu) = \ord_t(1-\nu) = 0$ 
   and $\ord_t(\nu) >0,\; \ord_t(1-\mu) > 0$ we have
   \begin{displaymath}
      i(\gamma,\Gamma_{F_1},{\mathbf a_2}) = 
      {\ord}_t(\mu(1-\mu)) = {\ord}_t(1-\mu)  \qquad\quad    
      i(\gamma',\Gamma_{F_1},{\mathbf a_3}) = 
      {\ord}_t(\nu(1-\nu)) = {\ord}_t(\nu)
   \end{displaymath}     
   while at the same time
   \begin{displaymath}
      {\ord}_t(1-\mu) + {\ord}_t(\nu) \,=\, 
      {\ord}_t(\mu(1-\mu)\nu(1-\nu))  \,=\, 
      {\ord}_t(F_1 F_2) \,=\, 0 \pmod{2} 
   \end{displaymath} 
   The last two equalities relate to substituting 
   $(x,y,z) \rightarrow (\mu(t),\nu(t),1)$ in $F_1\,,F_2$, and using 
   eq.~(\ref{eq:106}) which is valid for arbitrary branch $\gamma$ of 
   $\Gamma_F$. It follows that 
   $i(\gamma,\Gamma_{F_1},\bfa_2) + i(\gamma',\Gamma_{F_1},\bfa_3)$ is 
   even for  arbitrary pair $(\gamma,\gamma')$, implying (\ref{eq:550}).  
\end{proof}

\end{appendix} 

\vfill\eject

\end{document}
\bye